\documentclass[12pt]{article}

\usepackage[dvips]{graphicx}
\renewcommand{\baselinestretch}{1.75}
\newcommand \be{\begin{equation}}
\newcommand \ee{\end{equation}}
\newcommand \ba{\begin{eqnarray}}
\newcommand \ea{\end{eqnarray}}
\begin{document}

\def\today{\ifcase\month\or
 January\or February\or March\or April\or May\or June\or
 July\or August\or September\or October\or November\or
December\fi
 \space\number\day, \number\year}
%

\hfil PostScript file created: \today{}; \ time \the\time \
minutes
\vskip .15in

\centerline {LIKELIHOOD ANALYSIS OF EARTHQUAKE FOCAL MECHANISM
DISTRIBUTIONS }

\vskip .15in
\begin{center}
{Yan Y. Kagan and David D. Jackson}
\end{center}
\centerline {Department of Earth and Space Sciences,
University of California,}
\centerline {Los Angeles, California 90095-1567, USA;}
\centerline {Emails: {\tt kagan@moho.ess.ucla.edu,
david.d.jackson@ucla.edu}}

\vskip 0.12 truein


\vskip .15in
\noindent
{\bf Short running title}:
{\sc
Likelihood Analysis of Focal Mechanisms
}

\vskip .15in

Corresponding author contact details:

Office address:
Dr. Y. Y. Kagan, Department Earth and Space Sciences (ESS),
Geology Bldg, 595 Charles E. Young Dr.,
University of California Los Angeles (UCLA),
Los Angeles, CA 90095-1567, USA

Office phone  310-206-5611

FAX:          310-825-2779

E-mail: kagan@moho.ess.ucla.edu

\vskip 0.25in

SECTION: SEISMOLOGY

\newpage

\noindent
{\bf SUMMARY}
\hfil\break
In our paper published earlier
we discussed forecasts of earthquake focal mechanism and ways
to test the forecast efficiency.
Several verification methods were proposed, but they were
based on ad-hoc, empirical assumptions, thus their
performance is questionable.
In this work we apply a conventional, likelihood
method to measure a skill of forecast.
The advantage of such an approach is that earthquake rate
prediction can in principle be adequately combined with focal
mechanism forecast, if both are based on the likelihood
scores, resulting in a general forecast optimization. 
To calculate the likelihood score we need to compare actual
forecasts or occurrences of predicted events with the null
hypothesis that the mechanism's 3-D orientation is random. 
For double-couple source orientation the random probability
distribution function is not uniform, which complicates the
calculation of the likelihood value.
To better understand the resulting complexities we calculate
the information (likelihood) score for two rotational
distributions (Cauchy and von Mises-Fisher), which are used to
approximate earthquake source orientation pattern.
We then calculate the likelihood score for earthquake source
forecasts and for their validation by future seismicity data.
Several issues need to be explored when analyzing
observational results: their dependence on forecast and data
resolution, internal dependence of scores on forecasted angle,
and random variability of likelihood scores.
In this work we propose a preliminary solution to these
complex problems, as these issues need to be explored by a
more extensive statistical analysis.

\vskip 0.05in
\noindent
{\bf Key words}:
\vskip .05in
Probabilistic forecasting;
Earthquake interaction, forecasting, and prediction;
Seismicity and tectonics;
Theoretical seismology;
Statistical seismology;
Dynamics: seismotectonics.

\newpage

\section{Introduction}
\label{intro}
Kagan \& Jackson (2014) discussed two problems: forecasting
earthquake focal mechanisms and evaluating a forecast skill.
The first problem was initially addressed by Kagan \& Jackson
(1994), but no attempt verifying the forecast has been carried
out until now.
Kagan \& Jackson (2014) proposed several verification methods,
but the techniques were based on ad-hoc, empirical
assumptions, thus their performance is not clear.
In this work we apply a conventional, likelihood method to
measure the skill of a forecast. 

The likelihood estimate for the focal mechanism prediction
compares actual forecasts or later occurrences of predicted
events with the null hypothesis that the mechanism's
orientation is random. 
It is similar to our forecast testing for long- or short-term
earthquake rate predictions (Kagan \& Jackson, 1994, 2011),
where we use a uniform Poisson process in space or time as
the null hypothesis.
With the earthquake source orientation distribution, the
situation is more complex, the probability distribution
function (PDF) is not uniform, though its analytical
expression is known: see Kagan (1990, eq. 3.1) or Table 3 for
orthorhombic symmetry by Grimmer (1979).

Kagan \& Jackson (1994, 2011) approximate earthquake rate
forecast by smoothing the past seismicity record with a
spatial kernel.
They optimize the kernel by searching for the best prediction
of future seismicity level (see also Molchan, 2012). 
The assumption is that the {\it true} model belongs to a
general class of parametric models which is used to
approximate seismicity. 
Right now we adjust only a width of the smoothing kernel and
its functional form.
In principle though several other kernel parameters (such as
its directivity, magnitude dependence, etc.) can be optimized
(see, for instance, Kagan \& Jackson, 2011).

To measure the forecast skill several likelihood scores are
calculated (Kagan, 2009, Table~3), most of these scores
($I_0$, $I_1$, $I_2$, and $I_3$) have similar values, their
difference is due to forecast map resolution and other known
and controlled factors. 
This feature is explained by our smoothing forecast procedure
which yields a relatively smooth seismicity map.

A major problem with the focal mechanism forecast is that we
lack a model for earthquake source pattern similar to that for
earthquake rate.
Thus our forecast distribution contains many relatively sharp
steps, and if integrated they produce the $I_4$ score (Kagan,
2009, Table~3), which yields a consistent estimate that
converges to the true value only when the sample size tends to
infinity (Kagan, 2007; Molchan, 2010).
Therefore in this work we are trying to find the properties of
$I_4$ score for the earthquake catalogs under investigation.
Nevertheless, we hope that this new measure of focal mechanism
forecast skill would be useful in earthquake prediction
efforts.

\section{Statistical focal mechanism forecasts }
\label{stat}

Kagan \& Jackson (2014) analyzed the focal mechanism
orientation distribution in the Global Centroid-Moment-Tensor
(GCMT) catalog (Ekstr\"om {\it et al.}\ 2012, and its
references, see Fig.~1 in Kagan \& Jackson, 2014).
Kagan \& Jackson (2014) considered $\Phi_1$ angle which is the
average rotation angle between the forecasted weighted focal
mechanisms and the mechanisms of the 1977-2007 earthquakes in
a 1000~km circle surrounding this predicted event.
The $\Phi_2$ angle is the angle between the forecasted
mechanism and the double-couple ($DC$) mechanism of the
observed events in the 2008-2012 period.

In this work we limited our study to the forecast in the
latitude bandwidth $[ 75^\circ S-75^\circ N ]$, because
calculations in this range take considerably less time.
Moreover, almost all earthquakes in the GCMT catalog are
concentrated there (Kagan \& Jackson, 2014, Fig.~2).

\begin{figure}
\begin{center}
\includegraphics[width=0.75\textwidth]{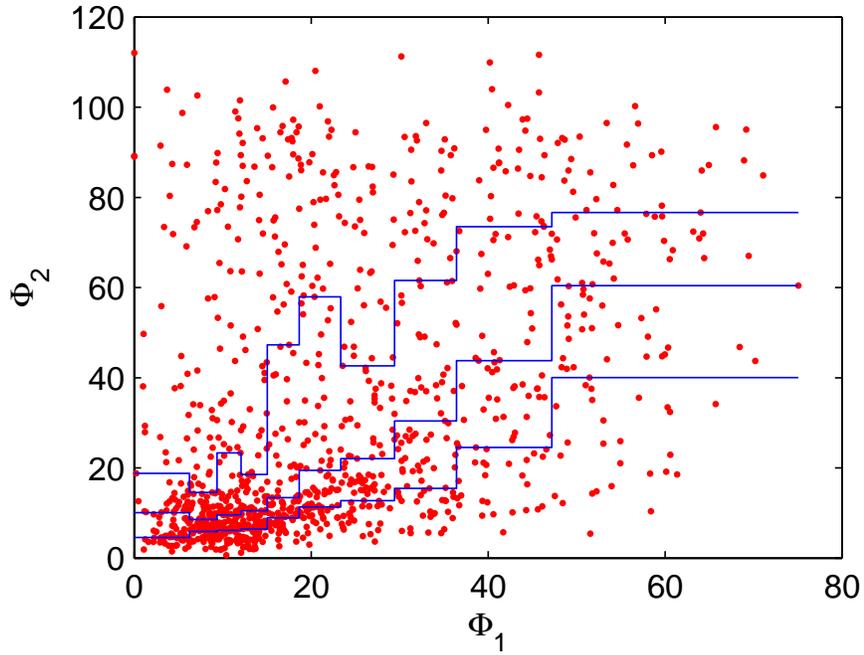}
\caption{Distribution of rotation angles }
\label{fig02}
\end{center}
\vskip -.5cm
in the Global Centroid Moment Tensor (GCMT) catalog,
1977--2007/2008--2012, latitude range $[ 75^\circ S-75^\circ N
]$, earthquake number $n=1069$.
Scatterplot of interdependence of the predicted $\Phi_1$ and
observed $\Phi_2$ angles.
Blue lines from top to bottom are 75\%, 50\%, and 25\%
quantiles, for a $\Phi_1$ angle subdivision with equal number
of events in 10 subsets. 
\end{figure}

In Fig.~\ref{fig02} we display a scatterplot of two angles
$\Phi_1$ and $\Phi_2$.
To investigate the interdependence of two angles, we subdivide
the plot of 1069 $\Phi_1$ angles into ten subsets with an
increasing angle and calculate the quantiles of $\Phi_2$
distribution in each of these subsets.

The focal mechanism forecast is displayed in Fig.~4 of Kagan
\& Jackson (2014).
About 42\% of the cells in the forecast have variables
$\Phi_1$ and $\Phi_2$ equal to zero.
These cell centers do not have any earthquake centroid within
1000~km distance.
To avoid future `surprises', in our earthquake rate forecasts
we assume that 1\% of all earthquakes occur uniformly over the
Globe (Jackson \& Kagan 1999; Kagan \& Jackson 2011).
We did not make such or similar an assumption for the focal
mechanism forecast, since it is not known what default value
needs to be adopted for these cells (Kagan \& Jackson, 2014).

However, almost all 2008-2012 earthquakes occurred in the
places within 1000~km of 1977-2007 epicentroids, thus their
angles could be evaluated.
Only 3 events out of 1069 are outside of the 1000~km limit.
In Fig.~\ref{fig02} these zero values of $\Phi_1$ and
$\Phi_2$ are all plotted at the point [0.0,~0.0] and thus are
not separately visible.
In our likelihood studies below we sometimes use only 1066
events for which the angles can be measured.

In Table~\ref{tab:Tab3} we list the properties of both angles
distribution shown in Fig.~\ref{fig02}.
The average $\Phi_2$ value increases steadily with the
increase of $\Phi_1$, though the standard deviation is
generally stable, thus the coefficient of variation also
decreases for later subsets.
In Fig.~7 by Kagan \& Jackson (2014) this interdependence of
two angles was characterized by regression lines.

\section{Rotation angle distributions }
\label{dcr1b}

Kagan (2013, Section~5) considers three statistical
distributions for the double-couple ($DC$) source orientation:
\hfil\break
$\bullet$
1. The uniform random rotation, which corresponds to
orthorhombic symmetry for a general $DC$ source.
This distribution is defined for the orientation angle range
$0^\circ - 120^\circ$.
The probability density function (PDF) is
\be
f(\Phi) \ = \ (4/\pi)(1 - \cos \Phi) \quad {\rm for} \quad
0 \le \Phi \le 90^\circ
\, ;
\label{eq1}
\ee
\be
f(\Phi) \ = \ (4/\pi) (3 \sin \Phi + 2 \cos \Phi - 2)
\quad {\rm for} \quad
90^\circ \le \Phi \le \Phi_S
\, ;
\label{eq2}
\ee
and
\ba
f(\Phi) \ = \ & (4/\pi) \, \Bigg \{ 3 \sin \Phi + 2 \cos \Phi
- 2 \, -
\nonumber \\
& (6/\pi) \bigg [ 2 \, \sin \Phi \, \arccos \left ({{1 + \cos
\Phi} \over {- 2 \cos \Phi}} \right )^{1/2} -
\nonumber \\
& ( 1 - \cos \Phi) \, \arccos {{1 + \cos \Phi}
\over {-2 \cos \Phi}} \, \bigg] \Bigg \}
\quad {\rm for} \quad
\Phi_S \le \Phi \le 120.0^\circ \, ,
\label{eq3}
\ea
where
\be
\Phi_S \ = \ 2 \arccos \, (3^{-1/2})
\ = \ \arccos \, \left (- \, {1 \over 3} \right ) \ \approx \
109.47^\circ \, .
\label{eq4}
\ee

\hfil\break
$\bullet$
2. Two non-uniform statistical distributions: the rotational
Cauchy law (Kagan 1992) and the rotational von Mises-Fisher
(VMF) law (Kagan 2000; Mardia \& Jupp 2000, pp.~289-292;
Morawiec 2004, pp.~88-89). 

The Cauchy law in the 3-D Euclidean space is scale-invariant
for relatively small angles and it has a power-law tail for
large angles. 
The {\sl rotational} Cauchy is defined on the 3-D hypersphere
of a normalized quaternion (Kagan 1982; 1990). 
The PDF of the rotational Cauchy distribution can be written
as 
\be
f(\Phi) \ = \ {{2} \over {\pi}} \Biggr [
{{\kappa A^2 (1 + A^2) } \over
{(\kappa^2 + A^2)^2}} \Biggl ]
\ = \
{{4\,\kappa\,\left[ 1 - \cos ({\Phi}) \right]
}\over {\pi \,{{\left[ 1 + {\kappa^2} + \left(
{\kappa^2} - 1 \right) \,\cos ({\Phi}) \right] }^2}}} \, ,
\quad {\rm for} \quad
0^\circ \le \Phi \le 180^\circ \, ,
\label{eq8}
\ee
where $A = \tan(\Phi/2)$.

Similarly to the rotational Cauchy law the von Mises-Fisher
distribution is a Gaussian-shaped function defined on the 3-D
hypersphere of a normalized quaternion. 
It is concentrated near the zero angle $\Phi)$.
This distribution can be implemented to model random errors in
determining focal mechanisms.

\begin{figure}
\begin{center}
\includegraphics[width=0.65\textwidth]{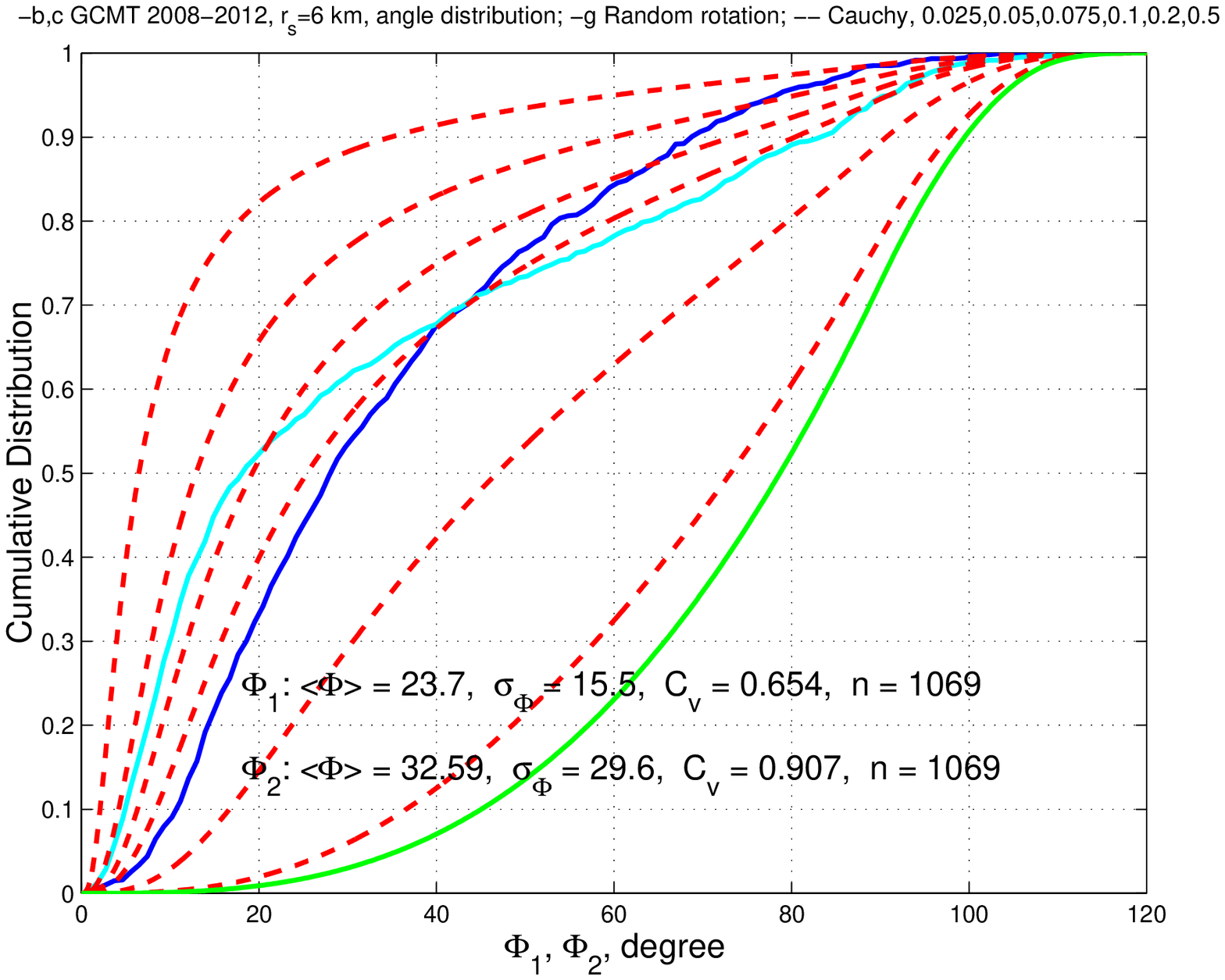}
\caption{GCMT catalog, 1977--2007, $r_s = 6$~km,
latitude range $[ 75^\circ S-75^\circ N ]$,
}
\label{fig05}
\end{center}
\vskip -.5cm
earthquake number $n=1069$.
Blue and cyan curves are cumulative distributions of predicted
rotation angle $\Phi_1$ and of observed rotation angle
$\Phi_2$ at earthquake centroids, respectively.
The dashed lines from left to right are for the Cauchy
rotational distribution with $\kappa = 0.025, 0.050, 0.075,
0.1, 0.2, 0.5$.
Right green solid line is for the random rotation.
Distribution analysis results for angles $\Phi_1$ and $\Phi_2$
are written at the bottom of the plot:
$<\Phi>$ is average rotation angle,
$\sigma_\Phi$ its standard deviation,
$C_v = { \sigma_\Phi \over <\Phi> } $ its coefficient of
variation.
\end{figure}

Fig.~\ref{fig05} displays the $DC$ random rotation
distribution (Eqs.~\ref{eq1}--\ref{eq4}), as well as several
Cauchy distributions. 
The Cauchy distribution (\ref{eq8}) has only one parameter
($\kappa$), a smaller $\kappa$-value corresponds to the
rotation angle $\Phi_{\rm min}$ concentrated closer to zero. 
These theoretical laws can be compared to the cumulative
$\Phi_1$ and $\Phi_2$ angle distributions.
The Cauchy law approximates the $\Phi_1$ curve reasonably well
for $\kappa = 0.075$ up to about 20$^\circ$.
Even for larger angles the observation curve is close to
the Cauchy law lines.
A similar effect is observed for the $\Phi_2$ curve.

The VMF distribution is obtained by generating a 3-D normally
distributed random variable {\bf u} ($u_1$, $u_2$, $u_3$) with
the standard deviation $\sigma_{\bf u}$ ($\sigma_{u_1}, \,
\sigma_{u_2}, \, \sigma_{u_3}$) and then calculating the unit
quaternion
\ba
q_0 \ = \ 1/ \sqrt { 1 + u_1^2 + u_2^2 + u_3^2 } \, , &&
\nonumber\\
q_i \ = \ u_i / \sqrt { 1 + u_1^2 + u_2^2 + u_3^2 } \, ,
\quad {\rm for} \ i = 1, 2, 3. &&
\label{eq10}
\ea
The 3-D rotation angle is calculated
\be
\Phi \ = \ 2 \arccos (q_0) \, .
\label{eq11}
\ee
Since components of the vector {\bf u} are normally
distributed, the sum ($u_1^2 + u_2^2 + u_3^2$) in
Eq.~\ref{eq10} follows the Maxwell distribution. 
The Maxwell equation describes the distribution of a vector
length in three dimensions, if the vector components have a
Gaussian distribution with a zero mean and a standard error
$\sigma_u$.
For small $\sigma_u$ the distribution of angle $\Phi$ follows
the Maxwell law (Kagan, 2013).

These Cauchy and VMF laws are theoretically defined for
orientation angle range $0^\circ - 180^\circ$.
However, because of the orthorombic symmetry of the general 
$DC$ source (Kagan, 2013) its maximum disorientation cannot
exceed $120^\circ$. 
Since we cannot obtain an analytical representation for these
distributions, we use simulation (Kagan, 1992) to derive the
distribution form for the range $0^\circ - 120^\circ$.

\begin{figure}
\begin{center}
\includegraphics[width=0.65\textwidth]{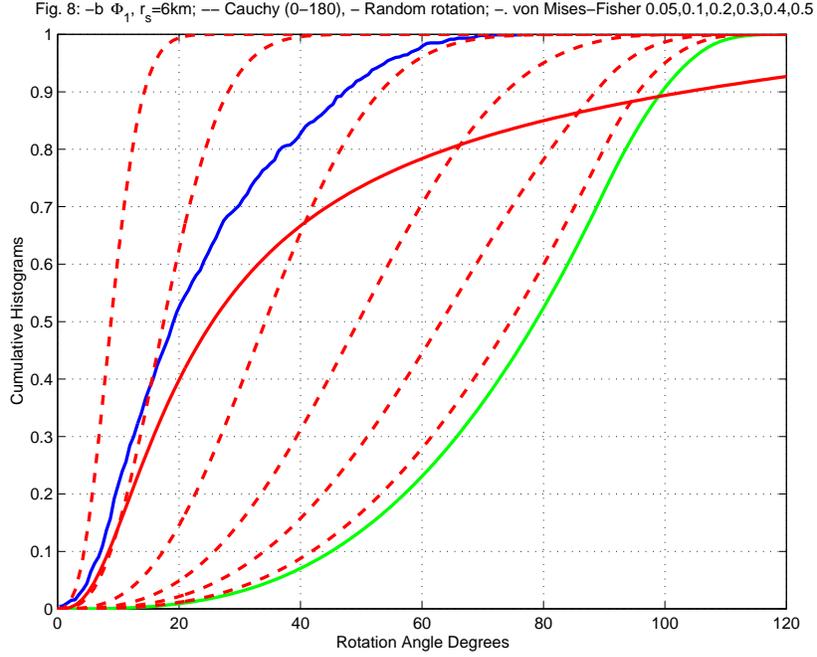}
\caption{GCMT catalog, $r_s = 6$~km, latitude range $[
75^\circ S-75^\circ N ]$,
}
\label{fig07}
\end{center}
\vskip -.5cm
1977--2007/2008--2012, earthquake number $n=1069$.
Blue curve is the cumulative distribution of predicted
rotation angle $\Phi_1$ at earthquake centroids. 
The dashed lines from left to right are for the von
Mises-Fisher (VMF) rotational distribution with $\sigma_u =
0.05, 0.1, 0.2, 0.3, 0.4, 0.5$.
Right green solid line is for the random rotation.
Red solid curve is for the rotational Cauchy distribution with
the domain $[ \, 0^\circ$ -- $180^\circ]$.
\end{figure}

In Fig.~\ref{fig07} the focal mechanism angles are compared to
the VMF distribution.
As expected (see Kagan, 2013) the fit of the VMF law to
observation curve is not as good as for the Cauchy
distribution, which provides a much better approximation
(Fig.~\ref{fig05}).

\section{Error diagrams }
\label{YYK_point2}

Table~\ref{tab:Tab1} shows the values of the information
scores, $I$, (Kagan, 2009) for two theoretical distributions.
As can be seen from Figs.~\ref{fig05} and \ref{fig07} the
curves that are close to the random rotation line have the
score value closer to zero.

There are considerable problems in computing the information
score for the rotation angles $\Phi_1$ and $\Phi_2$ in the
GCMT catalog.
As Kagan (2009) discusses, we have a reasonable model for the
earthquake rate forecast: future earthquakes are concentrated
close to the location of the past events, thus our general
forecast model is obtained by smoothing the past earthquake
locations to infer the spatial distribution for future events.
We evaluate the smoothing kernel parameters using some optimal
criteria (Kagan \& Jackson, 2011). 
The error or concentration diagrams for the forecast maps are
relatively smooth so that various methods for evaluating the
scores ($I_0$, $I_1$, $I_2$, and $I_3$) yield similar results
Kagan (2009, Table~3). 

However, no such general model is yet available for the
distributions of angles $\Phi_1$ and $\Phi_2$. 
Since the number of observations is relatively small, the
concentration diagrams or cumulative distributions for these
angles contains step-like jumps. 
Thus, we can calculate the score which Kagan (2009) called
$I_4$, the estimate of which is biased for small samples
(Kagan, 2007).
\be
I_4 \, = \, { 1 \over n } \sum_{i=1}^{n} \nu_i \,
\log_2 \left [ { { \nu_i }
/ \sum_{k_{i-1}}^{k_i} {\tau_i } } \right ]
\, ,
\label{Eq_inf6}
\ee
where $\tau$ is the cumulative fraction of the alarm time,
$\nu$ is the cumulative fraction of failures to predict, and
$k_i$ is the cell number corresponding to the $i$-th event,
$\log_2$ is used to obtain the score measured in the Shannon
bits of information (Kagan, 2009). 

In Table~\ref{tab:Tab2} several estimates of the $I_4$ score
are shown for both angles and for two choices of smoothing
kernel width ($r_s$).
We subdivide the angle range ($0^\circ - 120^\circ$) into
various grid cells to see how it influences the $I_4$-value.
The number of cells with non-zero number of events is
relatively small for large cell size, but for a finer
subdivision it approaches the total number of angle
measurements (1066).
The score value also approaches an upper limit for a finer
subdivision.
The results do not appear to depend on the $r_s$-value.

The final $I_4$-values can be reasonably well forecasted by
comparing their approximation by the Cauchy distribution in
Fig.~\ref{fig05} with the appropriate score values in
Table~\ref{tab:Tab1}.
For example, the major part of the $\Phi_1$ curve is between
Cauchy curves $\kappa=0.05$ and $\kappa=0.075$, and the score
value is also between the respective values in
Table~\ref{tab:Tab1}.

Table~\ref{tab:Tab4} shows the values of the information
scores for the 2008-2012 GCMT catalog in a format similar to
Table~3 by Kagan (2009).
We vary several parameters to investigate the dependence of
scores.
For example, the scores change with the grid modification,
partly because almost all the events are located in separate
cells for a higher-resolution forecast.
However, since the score $I_2$ is calculated for the actual
earthquake location in the test period, their value does not
depend on the cell size, as expected.

The influence of the smoothing kernel width ($r_s$) on the
score looks insignificant. 
More study is necessary to optimize our forecast by changing
($r_s$); unfortunately the needed computations are very
extensive.

\begin{figure}
\begin{center}
\includegraphics[width=0.65\textwidth]{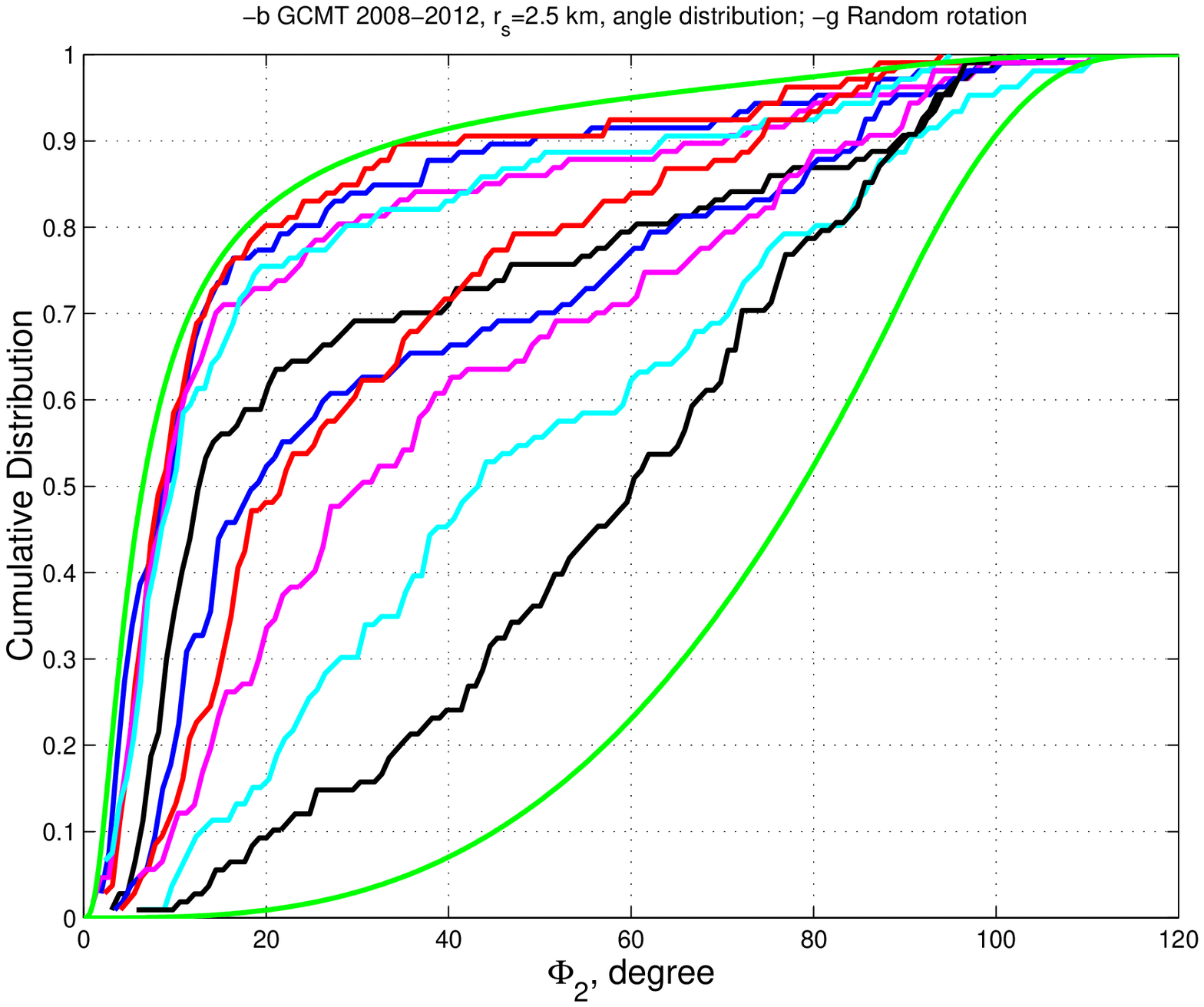}
\caption{GCMT catalog, 2008--2012, $r_s = 2.5$~km,
latitude range $[ 75^\circ S-75^\circ N ]$,
}
\label{fig08}
\end{center}
\vskip -.5cm
earthquake number $n=1066$.
Curves are cumulative distributions of the observed rotation
angle $\Phi_2$ at earthquake centroids, for 10 subsets (see
Fig.~\ref{fig02} and Table~\ref{tab:Tab3}). 
Right green solid line is for the random rotation, left
green solid curve is for the rotational Cauchy distribution
with $\kappa = 0.025$.
\end{figure}

Fig.~\ref{fig08} displays cumulative $\Phi_2$ distributions
for 10 subsets shown in Fig.~\ref{fig02} and
Table~\ref{tab:Tab3}.
The distributions move from left to right; the distribution
for the first subset is close to the Cauchy $\kappa=0.025$ law
curve, whereas the last distribution is close to the random
curve.
The score values shown in the last column of
Table~\ref{tab:Tab3} confirm this pattern: for the initial
subsets $I_4$-value is close to that $\kappa=0.25$ Cauchy
distribution (see Table~\ref{tab:Tab1}), whereas for the 10-th
subset $I_4$-value approaches zero.

\section{Discussion }
\label{disc}
The advantage of the likelihood approach for focal mechanism
orientation is that the likelihood scores for earthquake rate
prediction can be adequately combined with the focal mechanism
forecast, resulting in a general earthquake forecast
optimization.

\begin{figure}
\begin{center}
\includegraphics[width=0.65\textwidth]{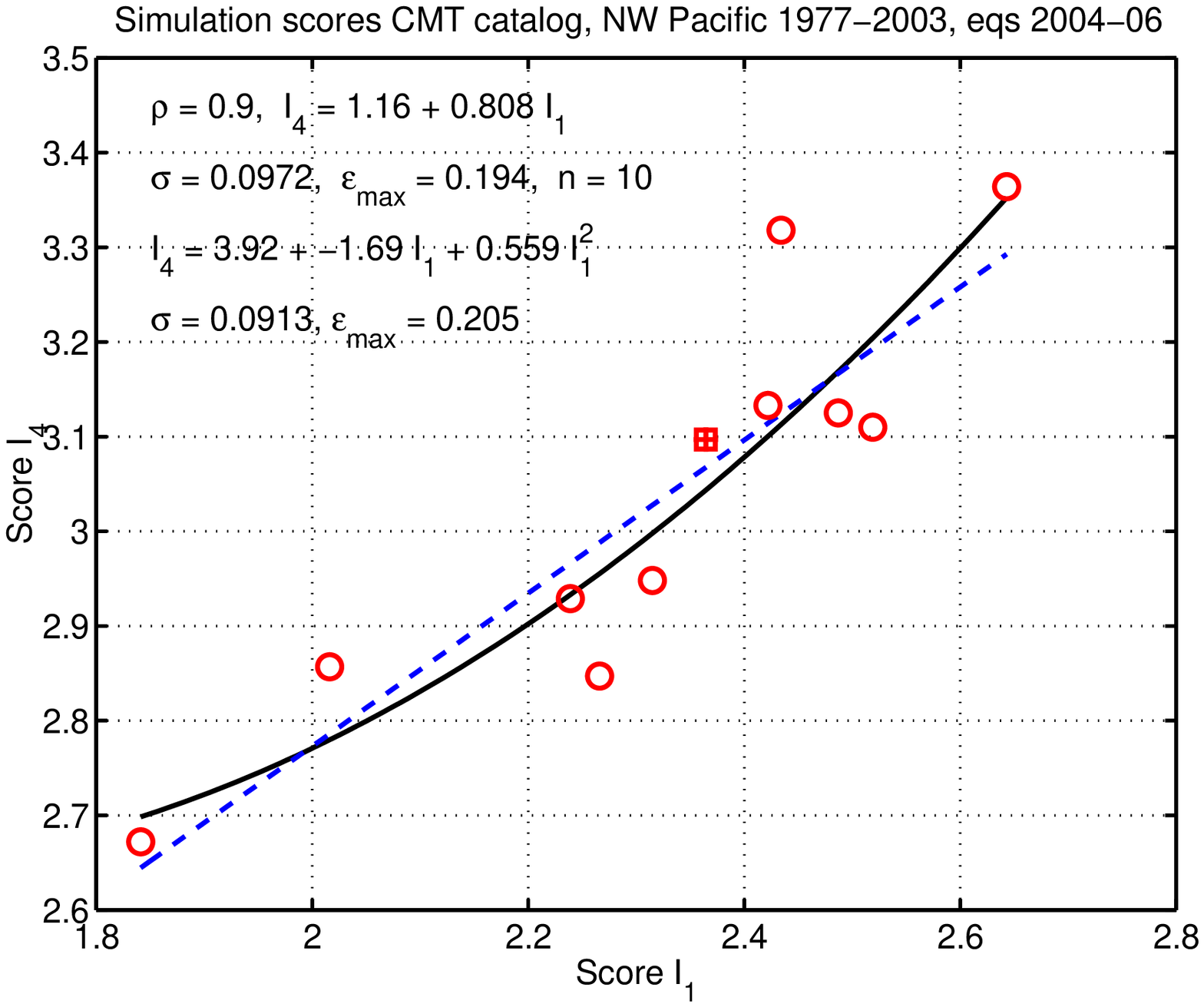}
\caption{GCMT catalog, 2004--2006, NW Pacific, $r_s = 15$~km,
}
\label{fig09}
\end{center}
\vskip -.5cm
earthquake number $n=108$.
Relation between two scores $I_1$ and $I_4$ in bits, shown by
circles, for first 10 simulated catalogs, displayed in Fig.~10
by Kagan (2009). 
We calculate two regression lines approximating the
interdependence of the $I_1$ and $I_4$ scores, the linear and
quadratic curves. 
The coefficient of correlation $\rho$ between the angles is
high $0.9$, indicating that the $I_4$ estimate can be well
evaluated by $I_1$ regression.
Results of both regressions are written at the top of the
plot: $\sigma$ -- standard error, $\varepsilon_{max}$ --
maximum difference.
A cross sign inside a square shows $I_1$ and $I_4$ estimates
for the NW Pacific region from Table~3 by Kagan (2009).
\end{figure}

As we observed (Kagan, 2009) the $I_4$ score estimates are
biased and have a higher random variation compared to the
other information scores.
In order to understand the properties of the $I_4$ score we
study a correlation between $I_4$ and other scores. To
investigate the relation between the two scores, $I_1$ and
$I_4$, in Fig.~\ref{fig09} we calculated these scores for 10
simulated catalogs shown in Fig.~10 (Kagan, 2009).
The range of $I_1$ values (0.8) can be compared with the
standard deviation for $I_0$ ($\sigma_n = 0.215$), shown in
Table~3 by Kagan (2009).
The scores $I_0$ and $I_1$ are optimized to have close values.
As the diagram demonstrates, $I_4$ is usually larger than
$I_1$, but their correlation coefficient is high, thus one can
estimate the $I_1$ score using regression.

As mentioned in the Introduction, a significant effort
needs to be extended to incorporate the methods developed in
this and the previous (Kagan \& Jackson, 2014) publications to
forecast earthquake focal mechanisms.
A similar investigation needs to be carried out to fully
optimize the global earthquake rate forecast (Kagan \&
Jackson, 2011).
However, these efforts will be mostly of technical nature,
the major forecast scientific issues are addressed in this and
above-mentioned papers.

\section{Conclusions }
\label{conc}
$\bullet$ \
1. We apply a likelihood method to measure the skill of an
earthquake focal mechanism forecast.
The advantage of such an approach is that the likelihood
scores for the earthquake rate prediction can quantitatively
be combined with the focal mechanism forecast, resulting in a
general forecast optimization.
\hfil\break
$\bullet$ \
2. We compare actual forecasts or occurrences of event source
properties with the null hypothesis that the mechanism's 3-D
orientation is random.
\hfil\break
$\bullet$ \
3. We calculate the information (likelihood) score for two
rotational distributions (Cauchy and von Mises-Fisher) which
are used to approximate a source orientation pattern.
\hfil\break
$\bullet$ \
4. We calculate the likelihood score for earthquake source
forecasts based on the GCMT catalog and their validation by
future seismicity data.
We explored the dependence of the results on data resolution,
internal dependence of scores on forecasted angle, and a
random variability of likelihood scores.

\subsection* {Acknowledgments
}
\label{Ackn}

The authors appreciate support from the National Science
Foundation through grants EAR-0711515, EAR-0944218, and
EAR-1045876, as well as from the Southern California
Earthquake Center (SCEC).
SCEC is funded by NSF Cooperative Agreement EAR-0529922 and
USGS Cooperative Agreement 07HQAG0008.
Publication 0000, SCEC.

\pagebreak

\centerline { {\sc References} }
\vskip 0.1in
\parskip 1pt
\parindent=1mm
\def\reference{\hangindent=22pt\hangafter=1}

\reference
Ekstr\"om, G., M. Nettles \& A.M. Dziewonski, 2012.
The global CMT project 2004-2010: Centroid-moment tensors for
13,017 earthquakes,
{\sl Phys.\ Earth Planet.\ Inter.}, {\bf 200-201}, 1-9.

\reference
Grimmer, H., 1979.
Distribution of disorientation angles if all relative
orientations of neighboring grains are equally probable,
{\sl Scripta Metallurgica}, {\bf 13}(2), 161-164,
DOI: 10.1016/0036-9748(79)90058-9.

\reference
Jackson, D.~D. \& Y.~Y.~Kagan, 1999.
Testable earthquake forecasts for 1999,
{\sl Seism.\ Res.\ Lett.}, {\bf 70}(4), 393-403.

\reference
Kagan, Y.~Y., 1991.
3-D rotation of double-couple earthquake sources,
{\sl Geophys.\ J. Int.}, {\bf 106}(3), 709-716.

Kagan, Y.~Y., 1992.
Correlations of earthquake focal mechanisms,
{\sl Geophys.\ J. Int.}, {\bf 110}(2), 305-320.

\reference
Kagan, Y. Y., 2000.
Temporal correlations of earthquake focal mechanisms,
{\sl Geophys.\ J. Int.}, {\bf 143}(3), 881-897.

\reference
Kagan, Y. Y., 2007.
On earthquake predictability measurement:
information score and error diagram,
{\sl Pure Appl.\ Geoph.}, {\bf 164}(10), 1947-1962.

\reference
Kagan, Y. Y., 2009.
Testing long-term earthquake forecasts:
likelihood methods and error diagrams,
{\sl Geophys.\ J. Int.}, {\bf 177}(2), 532-542.

\reference
Kagan, Y. Y., 2013.
Double-couple earthquake source: symmetry and rotation,
{\sl Geophys.\ J. Int.}, {\bf 194}(2), 1167-1179,
doi: 10.1093/gji/ggt156

\reference
Kagan, Y.~Y. \& D.~D.~Jackson, 1994.
Long-term probabilistic forecasting of earthquakes,
{\sl J. Geophys.\ Res.}, {\bf 99}, 13,685-13,700.

\reference
Kagan, Y. Y. \& D. D. Jackson, 2011.
Global earthquake forecasts,
{\sl Geophys.\ J. Int.}, {\bf 184}(2), 759-776.

\reference
Kagan, Y. Y. \& D. D. Jackson, 2014.
Statistical earthquake focal mechanism forecasts,
{\sl Geophys.\ J. Int.}, {\bf 197}(1), 620-629.

\reference
Mardia, K.\ V. \& P. E. Jupp 2000.
{\sl Directional Statistics},
Chichester, New York, Wiley, 429~pp.

\reference
Molchan, G., 2010.
Space-time earthquake prediction: the error diagrams,
{\sl Pure Appl.\ Geoph.}, ({\sl The Frank Evison Volume}),
{\bf 167}(8/9), 907-917. doi: 10.1007/s00024-010-0087-z.

\reference
Molchan, G., 2012.
On the Testing of Seismicity Models,
{\sl Acta Geophysica}, {\bf 60}(3), 624-637, doi:
10.2478/s11600-011-0042-0.

\reference
Morawiec, A.\ 2004.
{\sl Orientations and Rotations: Computations in
Crystallographic Textures},
Springer/Berlin, New York, pp.~200.


\newpage

\begin{table}
\begin{center}
\begin{tabular}{|r|r|c|c|r|c|c|}\hline
No. & $\Phi_1 $ & $< \Phi_2 >$ & $\sigma _\Phi$ & $C_v$ & n & $I_4$ \\
& $^\circ$ & $^\circ$ & $^\circ$ & Coef. & & Score
\\\hline
1 & \multicolumn{1} {|c|} 2 & \multicolumn{1} {|c|} 3 &
\multicolumn{1} {|c|} 4 & \multicolumn{1} {|c|} 5 & 6 & 7
\\\hline
1 & 0.2 & 18.0 & 23.3 & 1.29 & 106 & 6.56 \\
2 & 6.4 & 17.3 & 21.6 & 1.25 & 106 & 6.47 \\
3 & 9.5 & 20.8 & 25.6 & 1.23 & 107 & 5.93 \\
4 & 12.1 & 20.7 & 24.6 & 1.19 & 106 & 5.85 \\
5 & 15.1 & 30.6 & 30.9 & 1.01 & 107 & 4.32 \\
6 & 18.8 & 34.5 & 29.7 & 0.86 & 107 & 3.35 \\
7 & 23.5 & 31.7 & 24.0 & 0.76 & 106 & 3.42 \\
8 & 29.4 & 40.4 & 28.3 & 0.70 & 107 & 2.36 \\
9 & 36.4 & 50.4 & 28.9 & 0.57 & 106 & 1.45 \\
10 & 47.2 & 58.4 & 24.4 & 0.42 & 108 & 0.77 \\\hline
\end{tabular}
\caption{
Values of observed rotation angle $\Phi_2$ for 10 subsets
of scatterplot of two angles $\Phi_1$ and $\Phi_2$ (see
Fig.~\ref{fig02}).
The beginning value of the $\Phi_1$ interval is shown in
column~2.
$< \Phi_2 >$ is the average angle, $\sigma _\Phi$ is the
standard deviation, $C_v$ is the coefficient of variation 
($C_v = { \sigma_\Phi \over <\Phi> } $) for angle $\Phi_2 $,
$n$ the number of events in a subset, and $I_4$ is the
information score in bits.
}
\label{tab:Tab3}
\end{center}
\end{table}

\newpage

\begin{table}
\begin{center}
\begin{tabular}{|r|c|r|c|r|}\hline
\# & Parameter & Score & Parameter & Score \\\hline
& \multicolumn {2} {|c|} {Cauchy ($\kappa$)}
& \multicolumn {2} {|c|} {VMF ($\sigma_u$)} \\
1 & 0.025 & 7.48 & 0.050 & 8.15 \\
2 & 0.050 & 4.86 & 0.100 & 5.21 \\
3 & 0.075 & 3.49 & 0.200 & 2.44 \\
4 & 0.100 & 2.60 & 0.300 & 1.03 \\
5 & 0.200 & 0.95 & 0.400 & 0.30 \\
6 & 0.500 & 0.05 & 0.500 & 0.03 \\\hline
\end{tabular}

\caption{
Information scores in bits for theoretical distributions
(see Section~\ref{dcr1b}).
$\kappa$ is the parameter of the rotational Cauchy
distribution (see Fig.~\ref{fig05}), $\sigma_u$ ditto for the
von Mises-Fisher (VMF) rotational distribution (see
Fig.~\ref{fig07}).
}
\label{tab:Tab1}
\end{center}
\end{table}

\newpage
\begin{table}
\begin{center}
\begin{tabular}{|r|r|r|r|r|c|r|c|r|c|}\hline
\# & Subdivision & $\xi$ & $\Phi_1$ & $\xi$ & $\Phi_2$ & $\xi$
& $\Phi_1$ & $\xi$ & $\Phi_2$ \\\hline
& & \multicolumn {4} {|c|} {$r_s=2.5$~km}
& \multicolumn {4} {|c|} {$r_s=6.0$~km}
\\\hline
1 & 120 & 70 & 4.18 & 108 & 3.49 & 69 & 4.04 & 108 & 3.48 \\
2 & 1200 & 448 & 4.41 & 538 & 3.74 & 452 & 4.28 & 531 & 3.70 \\
3 & 12000 & 955 & 4.74 & 975 & 4.03 & 957 & 4.59 & 952 & 4.03 \\
4 & 120000 & 1058 & 4.94 & 1053 & 4.18 & 1049 & 4.76 & 1057 &
4.23 \\
5 & 1200000 & 1064 & 4.97 & 1065 & 4.24 & 1050 & 4.77 & 1057
& 4.23 \\\hline
\end{tabular}

\caption{
Information scores in bits for $\Phi_1$ and $\Phi_2$ angles
(see Section~\ref{intro}) in the GCMT catalog,
$r_s$ is the width of the smoothing kernel, $\xi$ is the total
number of non-zero intervals (out of 1066 possible).
}
\label{tab:Tab2}
\end{center}
\end{table}

\newpage

\begin{table}
\begin{center}
\begin{tabular}{|r|c|r|r|r|}\hline
\# & & \multicolumn {3} {|c|}
{2008-2012, $[ 75^\circ S-75^\circ N ]$ }
\\ \cline {2-5}
  & Grid & $0.1^\circ$ & $0.5^\circ$ & $0.5^\circ$ \\\hline
  & $r_s$ & 2.5~km & 2.5~km & 6.0~km \\\hline
  & $\xi$ & 1025 & 758 & 758 \\\hline
 & Score & & & \\
1 & $I_0$ & 4.30 & 4.25 & 3.73 \\
2 & $I_1$ & 4.01 & 3.84 & 3.96 \\
3 & $I_2$ & 3.98 & 3.98 & 4.07 \\
4 & $I_3$ & 4.29 & 4.25 & 3.73 \\
5 & $I_4$ & 4.99 & 4.85 & 4.86 \\\hline
\end{tabular}
\caption{Information scores in bits for one event, $r_s$ is
the width of the smoothing kernel, $\xi$ is the total number
of non-zero intervals (out of 1069 possible). 
}
\label{tab:Tab4}
\end{center}
\end{table}

\clearpage

\newpage

\renewcommand{\baselinestretch}{1.75}

\parindent=0mm

\newpage

\end{document}